\documentclass[Physsubmission, Phys]{SciPost}
\hypersetup{
    colorlinks,
    linkcolor={red!50!black},
    citecolor={blue!50!black},
    urlcolor={blue!80!black}
}

\usepackage[bitstream-charter]{mathdesign}
\DeclareSymbolFont{usualmathcal}{OMS}{cmsy}{m}{n}
\DeclareSymbolFontAlphabet{\mathcal}{usualmathcal}

\begin{document}

\begin{center}{\Large \textbf{
Correlations between azimuthal asymmetries and multiplicity and mean transverse momentum in small collisions systems in the CGC
}}\end{center}

% TODO: write the author list here. Use initials + surname format.
% Separate subsequent authors by a comma, omit comma at the end of the list.
% Mark the corresponding author with a superscript *.
\begin{center}

Tolga Altinoluk\textsuperscript{1}, N\'estor Armesto\textsuperscript{2$\star$}, Alex Kovner\textsuperscript{3}, Michael Lublinsky\textsuperscript{4}, and Vladimir V. Skokov\textsuperscript{5,6}

\end{center}

\begin{center}
{\bf 1} National Centre for Nuclear Research, 02-093 Warsaw, Poland \\
{\bf 2} Departamento de F\'{i}sica de Part\'{i}culas and IGFAE, Universidade de Santiago de Compostela, 15782 Santiago de Compostela, Galicia-Spain \\
{\bf 3} Physics Department, University of Connecticut, 2152 Hillside Road, Storrs, CT 06269, USA \\
{\bf 4} Physics Department, Ben-Gurion University of the Negev, Beer Sheva 84105, Israel  \\
{\bf 5} North Carolina State University, Raleigh, NC 27695, USA\\
{\bf 6} RIKEN-BNL Research Center, Brookhaven National Laboratory, Upton, NY 11973, USA\\

* nestor.armesto@usc.es
\end{center}

\begin{center}
\today
\end{center}

% For convenience during refereeing (optional),
% you can turn on line numbers by uncommenting the next line:
%\linenumbers

\definecolor{palegray}{gray}{0.95}
\begin{center}
\colorbox{palegray}{
  \begin{tabular}{rr}
  \begin{minipage}{0.1\textwidth}
    \includegraphics[width=22mm]{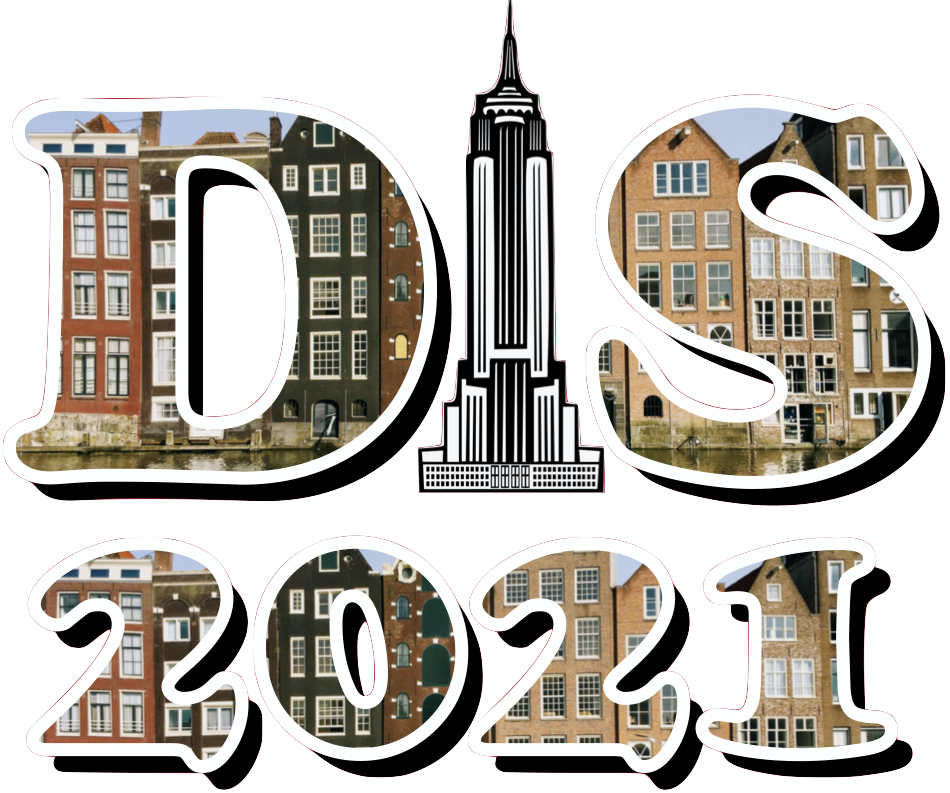}
  \end{minipage}
  &
  \begin{minipage}{0.75\textwidth}
    \begin{center}
    {\it Proceedings for the XXVIII International Workshop\\ on Deep-Inelastic Scattering and
Related Subjects,}\\
    {\it Stony Brook University, New York, USA, 12-16 April 2021} \\
    \doi{10.21468/SciPostPhysProc.?}\\
    \end{center}
  \end{minipage}
\end{tabular}
}
\end{center}

\section*{Abstract}
{\bf
% TODO: write your abstract here.
Considering a dilute-dense situation suitable for pA collisions, we compute in the Color Glass Condensate the correlation between azimuthal asymmetries, specifically the squa\-red second Fourier coefficient $v_2^2$, and the total multiplicity in the event. We also analyse the correlation between $v_2^2$ and the mean squared transverse momentum of particles in the event. In both cases, we find that the correlations are generally very small, consistent with the observations. We also note an interesting sharp change in the value of $v_2^2$ and its correlations as a function of the width of the transverse momentum bin, related with a change of the dominance of Bose and HBT quantum correlations.
}

%\vspace{10pt}
%\noindent\rule{\textwidth}{1pt}
%\tableofcontents\thispagestyle{fancy}
%\noindent\rule{\textwidth}{1pt}
%\vspace{10pt}

\section{Introduction}
\label{sec:intro}
% TODO: write your article here.
The measurement at the LHC and RHIC of many observables that behave similarly in small collision systems, pp and pA, and in heavy ion collisions, named the small system problem, is one of the key recent findings on the strong interaction, see~\cite{Nagle:2018nvi} and refs. therein. Among these features, the most celebrated one is the ridge: a maximum in two particle correlations elongated along pseudorapidity and peaked at zero and $\pi$ azimuthal angle. While in heavy ion collisions such azimuthal asymmetries are taken as signatures of a collective expansion describable by relativistic viscous hydrodynamics, in small systems this final state explanation looks tenuous and initial state ones have been essayed, see~\cite{Altinoluk:2020wpf} and refs. therein.

Concerning correlations between azimuthal asymmetries and global characteristics of the event, the ridge in small collision systems seems to be almost independent of the multiplicity in the collision~\cite{ATLAS:2017rtr}. Furthermore, the correlation of azimuthal asymmetries with the average transverse momentum has been proposed as sensitive to the initial geometry in the collision and to the initial or final dynamics underlying the correlations~\cite{Bozek:2016yoj,Giacalone:2020byk,Lim:2021auv}, and measured in~\cite{ATLAS:2019pvn}.

The aim of this contribution is to present and discuss the results on such correlations obtained in the framework of the initial state explanations provided by the CGC, see~\cite{Altinoluk:2020wpf}. In the CGC correlations come from the Bose enhancement of gluons in the wave function of the colliding hadrons and the HBT correlations of finally produced gluons~\cite{Altinoluk:2015uaa,Altinoluk:2015eka}. The work is based in the formalism developed in~\cite{Altinoluk:2018ogz} to compute two and three gluon correlations in pA collisions. Here we show and discuss the main results, referring the reader to~\cite{Altinoluk:2020psk} for full details.

\section{$v_2^2$ and correlations}
\label{sec:v2corr}

In the CGC, the number of produced gluons reads
\begin{equation}
\left.\frac{dN}{d^{2}kdy}\right|_{\rho_{p},\rho_{ t}}=\frac{2g^{2}}{(2\pi)^{3}}\int\frac{d^{2}q}{(2\pi)^{2}}\frac{d^{2}q'}{(2\pi)^{2}}\Gamma({k},{q},{q}')\rho_{p}^{a}(-{q}')\left[U^{\dagger}({k}-{q}')U({k}-{q})\right]_{ab}\rho_{ p}^{b}({q}),
\end{equation}
with
\begin{equation}
L^i(k,q)=\bigg[ \frac{(k-q)^i}{(k-q)^2}-\frac{k^i}{k^2}\bigg] ,\ \ 
\Gamma({k},{q},{q}')=
L(k,k-q)\cdot L(k,k-q'),
\end{equation}
$\rho_{\rm p}(p)$ a given configuration of the color charged density in the projectile, and $U(q)$ the eikonal scattering matrix -- adjoint Wilson line -- for scattering of a single gluon on a fixed configuration of target fields.
The target Wilson lines implicitly depend on the target color sources, $\rho_{\rm t}$.

Single inclusive and double inclusive gluon production are computed as
\begin{equation}
	\frac{dN^{(1)}}{d^{2}kdy}=\left\langle \left\langle  \left.\frac{dN}{d^{2}kdy}\right|_{\rho_{ p},\rho_{ t}} 
	\right \rangle_{ p} 
	\right\rangle_{ t} ,\ \ 
\frac{d N^{(2)}}{d^{2}k_{1}dy_{1}d^{2}k_{2}dy_{2}}= 
\left\langle \left \langle
\left. \frac{dN}{d^{2}k_{1}dy_{1}}\right|_{\rho_{ p},\rho_{ t}}
\left.\frac{dN}{d^{2}k_{2}dy_{2}} \right|_{\rho_{ p},\rho_{t}}
 \right\rangle_{ p} 
 \right\rangle_{ t} \, ,
 \label{eq:inclu}
\end{equation}
where the averaging is performed over the projectile and target color charge configurations.

The total multiplicity $N$ per unit of rapidity and mean transverse momentum squared per particle $\overline{k^2}$ are calculated as ($\int_\phi\equiv \int d\phi,\int_k\equiv\int d^2k$)
\begin{equation} \label{barN}
	 N\,=\, \int_k \frac{dN^{(1)}}{d^{2}kdy}\ ,\ \ 	\overline{k^2}\,=\, \frac{1}{  N}\int_k k^2\,\frac{dN^{(1)}}{d^{2}kdy}\ ,
	\end{equation}
and the azimuthal  flow harmonics $v^2_n$ defined as 
\begin{equation}
	%_
v^2_n(k_1,k_2)\equiv	   \int_{\phi_1, \phi_2} e^{i n(\phi_1-\phi_2)}   \frac { d^2N^{(2)} } { d^2k_1 d^2 k_2 }  \ \Bigg/ \int_{\phi_1, \phi_2}    \frac {  d^2N^{(2)} } { d^2k_1 d^2 k_2 }  \ ,
\end{equation}
where $\phi_1$, $\phi_2$ are the azimuthal angles of the corresponding transverse momenta. Below, we focus on $v^2_2$ only.
In this framework, each collision event corresponds to a fixed configuration of $\rho_p$ and $\rho_t$, with
the averaging  introduced in (\ref{eq:inclu}) being equivalent to averaging over all possible events. 

Studying the dependence of any observable on multiplicity would require to select from the total ensemble only events  with  total multiplicity in some multiplicity bin  and calculate the observable by averaging only over those events. In practice this has not yet been accomplished, so we choose to compute the correlation between $v^2_2$ and  $N$,  i.e. $\langle\langle v^2_2(k_1,k_2)|_{\rho_p,\rho_t}N|_{\rho_p,\rho_t}\rangle_p\rangle_t$, and similarly between $v^2_2$ and the squared transverse momentum per particle. The averaging in these expressions goes over the whole ensemble of events, and thus there is no need to consider particular sub ensembles. For that we make use of the results for two and three gluon inclusive  production in pA collisions obtained in~\cite{Altinoluk:2018ogz}.

Our calculation is done using the MV model~\cite{McLerran:1993ni,McLerran:1993ka} for projectile ensembles that we consider translationally invariant, and taking only leading contributions in the number of sources (i.e., $Q_s^2S_\perp$, the squared saturation scale times the overlap area of the collision)~\cite{Kovner:2017ssr,Kovner:2018vec}, see also the talk by P. Agostini in this workshop. We work at leading order in the number of colours and use the GW model for the target average of two Wilson lines~\cite{Golec-Biernat:1998zce}. In order to proceed analytically as far a possible, we compute the results at leading power of $Q_s^2/k_i^2$, with $k_i$ the transverse momenta of the measured final gluons, thus only valid for $Q_s^2/k_i^2\ll 1$. Besides, we take only the leading correlated pieces in the two and three gluon inclusive cross section, denoted below by $Q$ and $X$ below, respectively. Finally, when required we use an IR cutoff $\lambda \sim 1/(Q_s^2S_\perp)\sim 1/25$ (the results show small sensitivity to the exact value of this cutoff).

The correlators that we compute are defined:
\begin{eqnarray}\label{ONv2}
%&&%{\cal O}_{ N,v_2}
&&\langle v^2_2(k_2,k_3)N\rangle\\
%=\int d\phi_2\, d\phi_3 \, e^{i2(\phi_2-\phi_3)}\int d^2k_1 \frac{dN^{(3)}}{d^2k_1\, d^2k_2\, d^2k_3}\ 
%\bigg/ \ 
% \int d\phi_2\, d\phi_3 \, e^{i2(\phi_2-\phi_3)} \frac{dN^{(2)}}{d^2k_2d^2k_3} \int d^2k_1 \frac{dN^{(1)}}{d^2k_1}\nonumber\\
&&=\int_{\phi_2, \phi_3} e^{i2(\phi_2-\phi_3)}\int_{k_1} \frac{dN^{(3)}}{d^2k_1\, d^2k_2\, d^2k_3}\bigg|_X \ 
\Bigg/ \ 
 \int_{\phi_2, \phi_3} e^{i2(\phi_2-\phi_3)} \frac{dN^{(2)}}{d^2k_2d^2k_3}\bigg|_Q \int_{k_1} \frac{dN^{(1)}}{d^2k_1}\, ,\nonumber 
\end{eqnarray}
\begin{eqnarray}\label{Okv2}
%{\cal O}_{ k,v_2}
&&\langle v^2_2(k_2,k_3)\overline{k^2}\rangle\\
&&=\int_{\phi_2, \phi_3} \, e^{i2(\phi_2-\phi_3)}\int_{k_1} \, k_1^2\, \frac{dN^{(3)}}{d^2k_1\, d^2k_2\, d^2k_3}\bigg|_X\ 
\Bigg/ \ 
 \int_{\phi_2, \phi_3}  e^{i2(\phi_2-\phi_3)} \frac{dN^{(2)}}{d^2k_2d^2k_3}\bigg|_Q Q_s^2\int_{k_1}\,  \frac{dN^{(1)}}{d^2k_1}\,,\nonumber
\end{eqnarray}
with additional integrals over $k_2$ and $k_3$ in regions $[k-\Delta/2,k+\Delta/2]$ and $[k^\prime-\Delta/2,k^\prime+\Delta/2]$, respectively, with $k,k^\prime\gg\Delta\sim Q_s$, both in numerators and denominators. Full explanations and detailed expressions can be found in~\cite{Altinoluk:2020psk}.

\section{Numerical results}
\label{sec:numres}

In all our results we plot momentum in units of $Q_s$ and the quantities of interest multiplied by $(N_c^2-1)S_\perp Q_s^2$ ($\sim 200$ for pPb collisions) in order to exhibit the universal features of the results applicable to any target (any $Q_s$) and projectile (any $S_\perp$). We explore the interplay between  the relative position of the centres of the two bins, $k$ and $k'$ and the bin width $\Delta$. $v^2_2$ receives contributions from two types of correlations: Bose and HBT. The width of the former in momentum space is naturally of order $Q_s$, while the latter have  much shorter range (we took them as delta functions). Thus we expect that when $|k- k'|<\Delta$ both the HBT and Bose effects will contribute to $v^2_2$. However, when there is no overlap between the two bins, the HBT correlation should disappear. We thus expect a steep variation when $k-k'\approx \Delta$.

\begin{figure}[htb!]
 \begin{center}
 \resizebox{0.45\textwidth}{!}{%
  \includegraphics{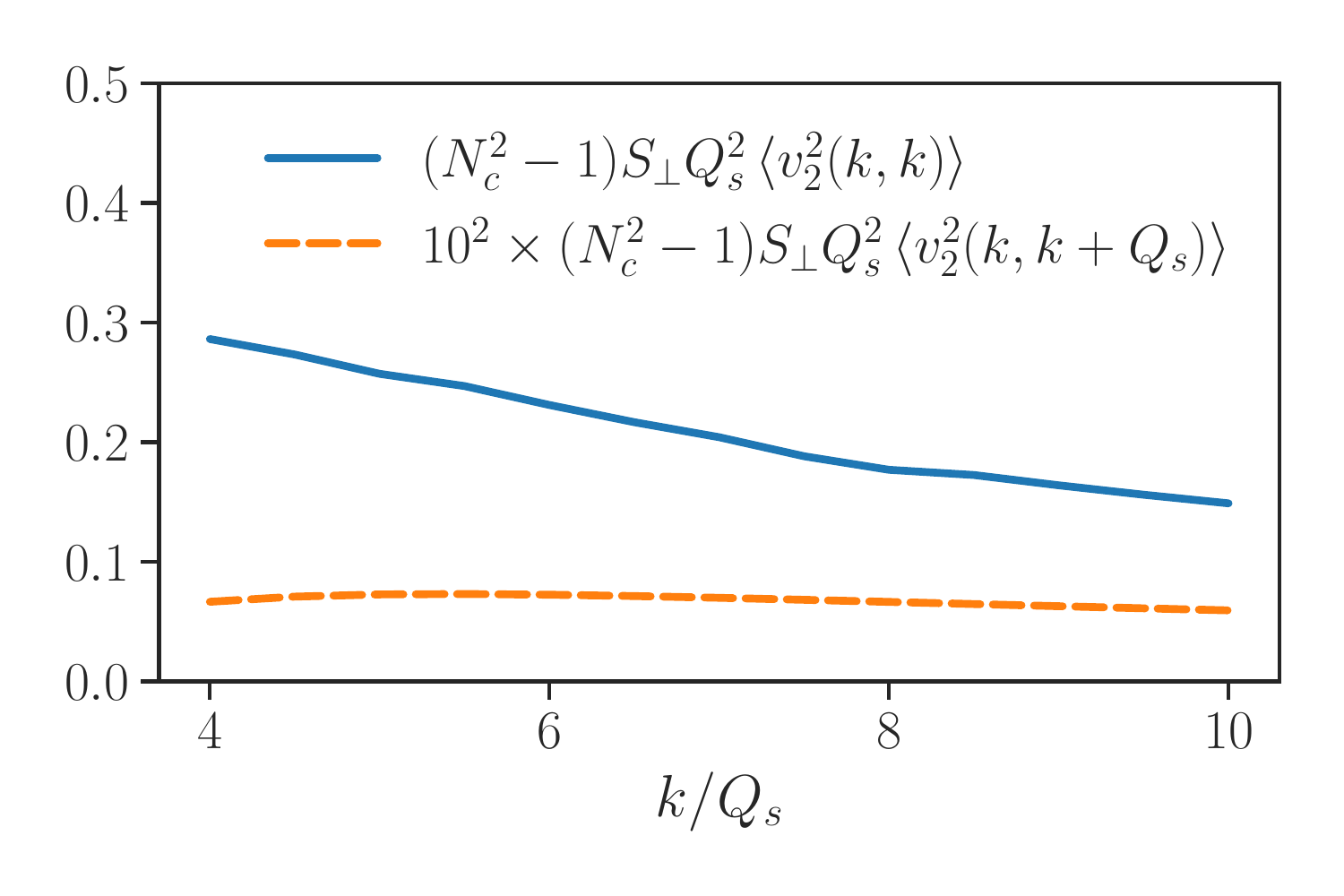}
}\hfill
\resizebox{0.45\textwidth}{!}{%
  \includegraphics{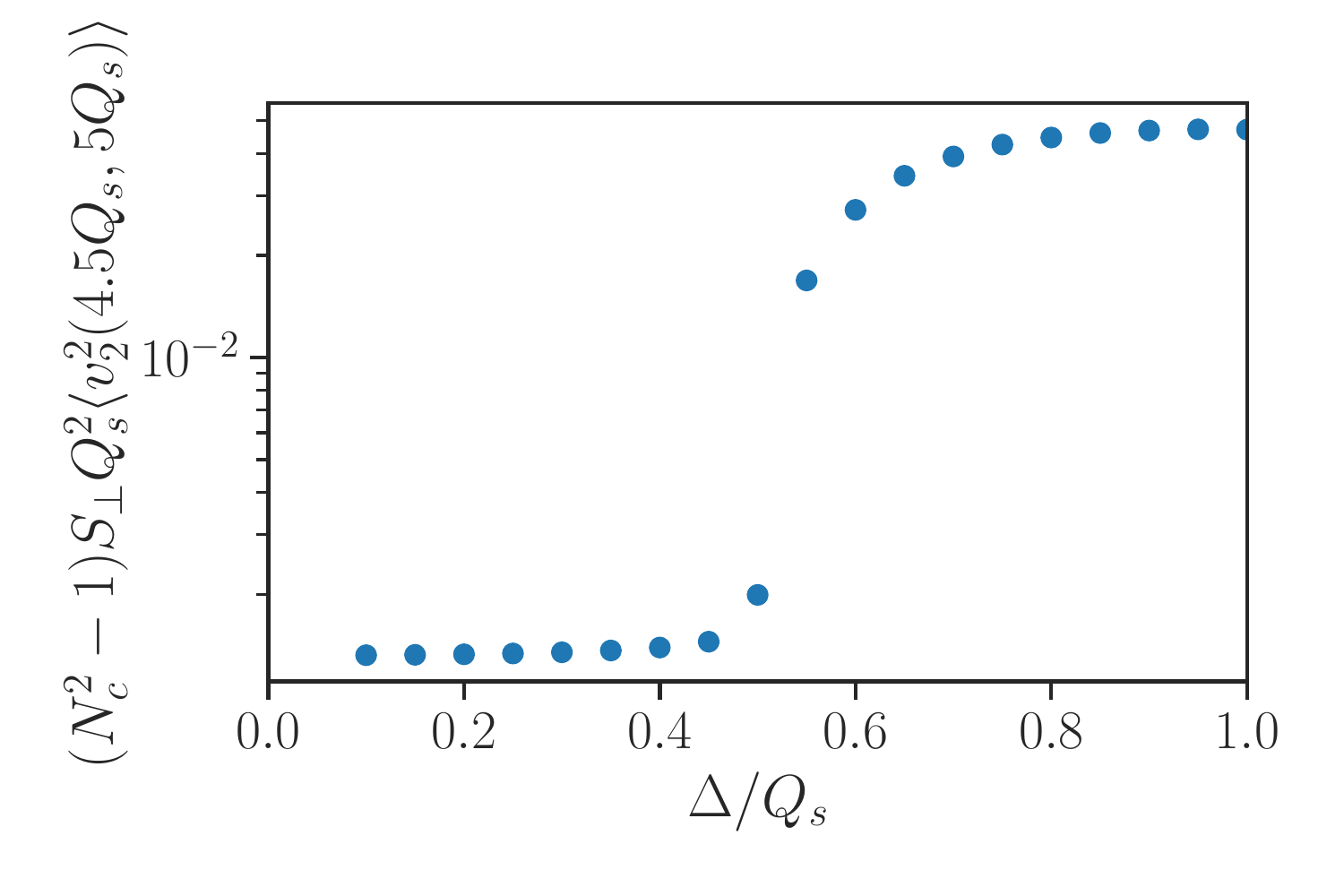}
}
\end{center}
\vskip -0.7cm
\caption{
Left  panel:
The second flow harmonic, $v_2^2$ as a function of transverse momentum. 
The calculation of $v^2_2$ is performed for two cases: a)  the same momentum of the pair, b) the momentum of the pair is offset by the saturation momentum of the target in order to avoid the gluon HBT effect. The bin width in both cases is $\Delta=Q_s/2$. 
Right panel:
	The second flow harmonic, $v_2^2$, as a function of the bin width.  The centres of the two bins are chosen at $k=4.5 Q_s$, $k'=5Q_s$.%Two particle correlations in $pp$ and $p$Pb collisions at the LHC measured by the ATLAS Collaboration~\cite{Aaboud:2016yar}, for different energies and particle multiplicities in the event.
}
\label{fig:0}       % Give a unique label
\end{figure}

Fig.~\ref{fig:0} shows our results for $v^2_2$.  The left panel shows the huge dominance of the HBT contribution, by a factor $\sim 50$, and the different transverse momentum dependence of Bose and HBT, while the right panel demonstrates the expected sharp change in $v^2_2$ at the point when the width of the interval equals the distance between the interval midpoints.

\begin{figure}[htb!]
\begin{center}
\resizebox{0.45\textwidth}{!}{%
  \includegraphics{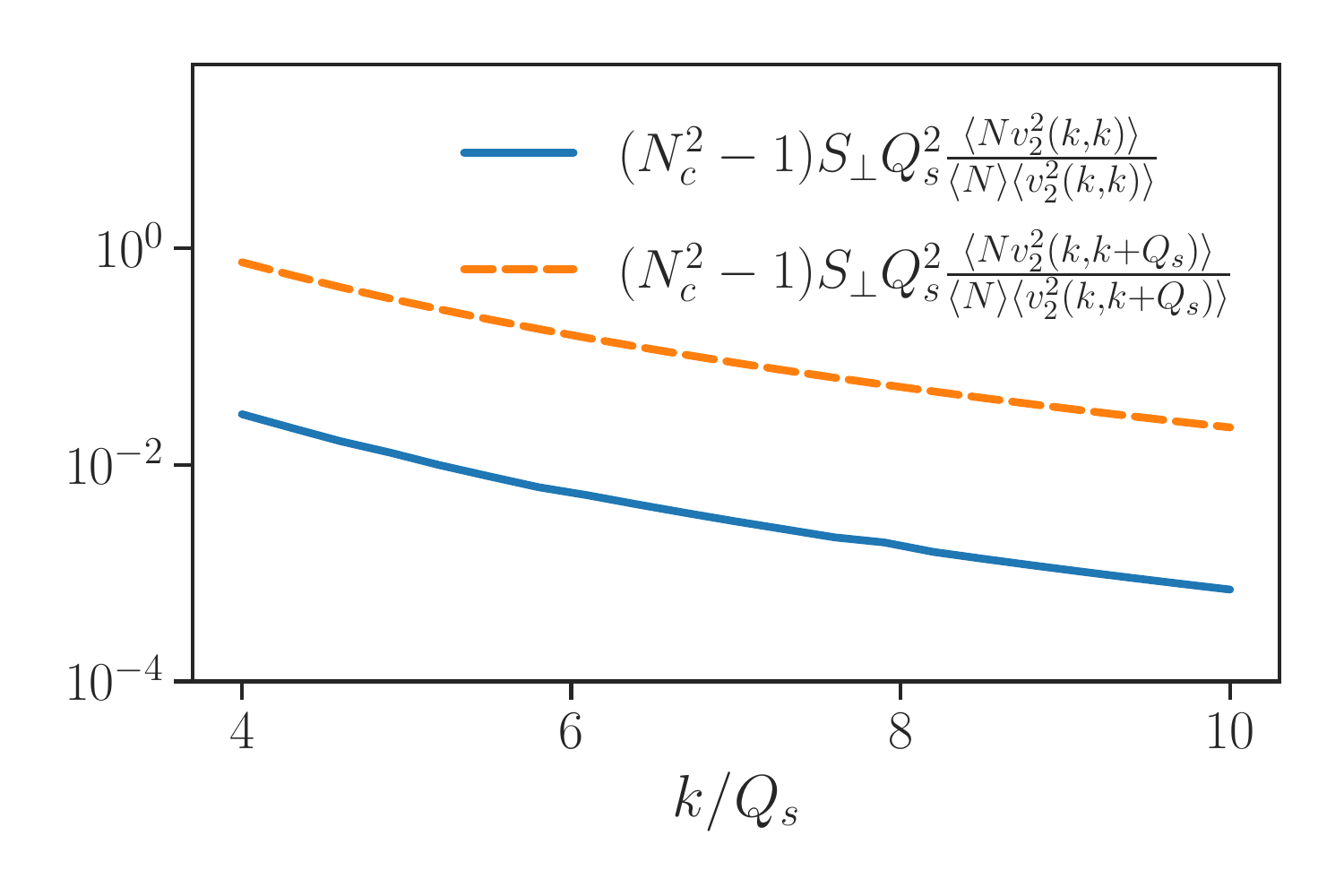}
}\hfill
\resizebox{0.45\textwidth}{!}{%
  \includegraphics{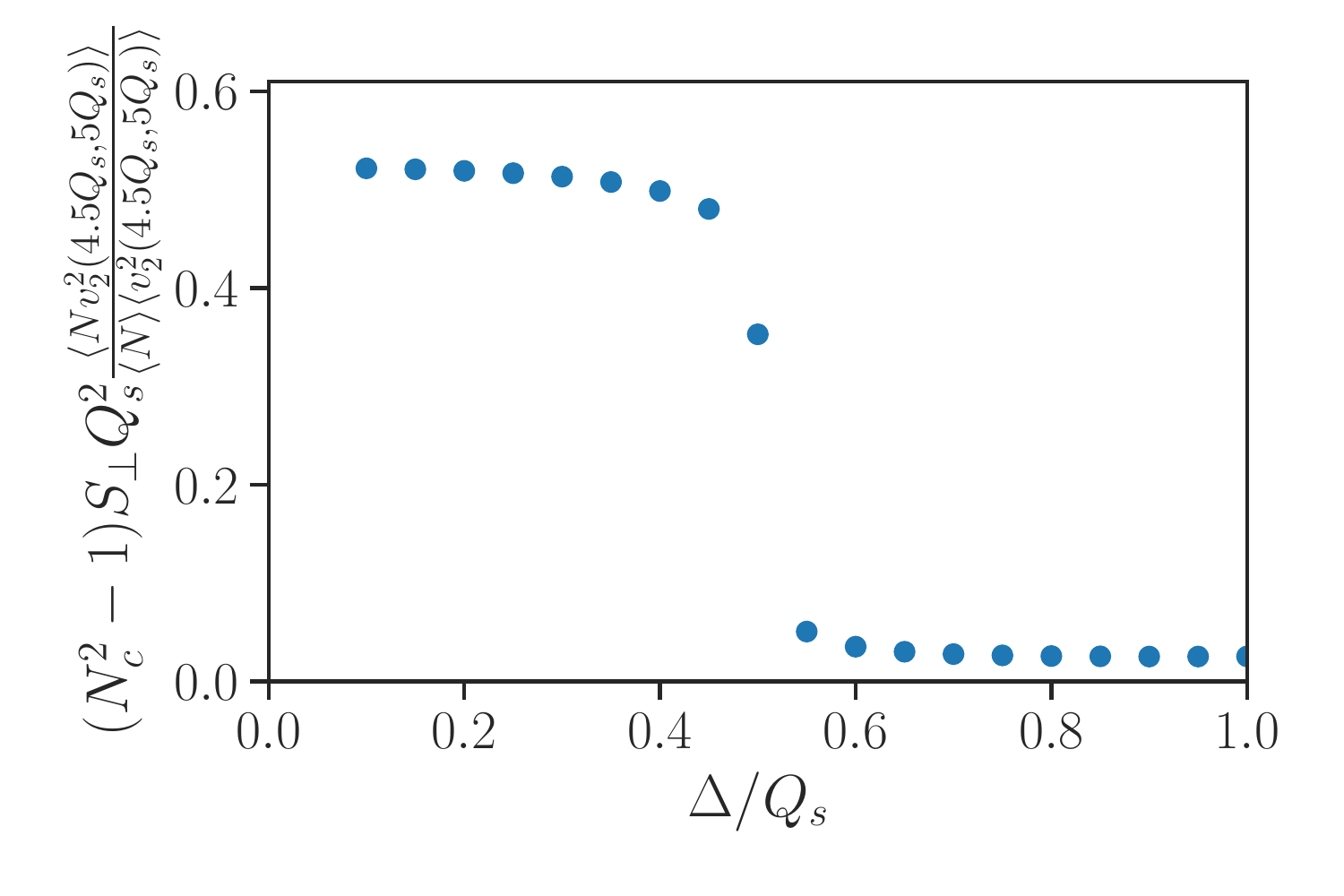}
}
\end{center}
\vskip -0.7cm
\caption{Id. to Fig.~\ref{fig:0} but for the three particle correlation function $\langle v^2_2N\rangle$ defined by  the normalised correlations between  $v^2_2$  and the total multiplicity of produced particles.
%Left  panel:
%	The three particle correlation function $\langle v^2_2N\rangle$ defined by  the normalised correlations between  $v^2_2$  and the total multiplicity of produced particles. The calculation of $v^2_2$ is performed for two cases: a)  the same momentum of the pair, b) the momentum of the pair is offset by the saturation momentum of the target in order to avoid the gluon HBT effect. The bin width in both cases is $\Delta=Q_s/2$. 
%%Two particle correlations in $pp$ and $p$Pb collisions at the LHC measured by the ATLAS Collaboration~\cite{Aaboud:2016yar}, for different energies and particle multiplicities in the event.
%Right panel: The three particle correlation function $\langle v^2_2N\rangle$ as a function of the bin width. 
}
\label{fig:1}       % Give a unique label
\end{figure}

Fig.~\ref{fig:1} shows our numerical results for the correlation function between $v^2_2$ and the total multiplicity (a similar behaviour is found for the correlation of $v^2_2$ with average squared transverse momentum).   
Fig.~\ref{fig:1} left shows that the normalised correlation function is strongly suppressed for values of bin width for which $v^2_2$ is sizeable, which is when the HBT effect in $v^2_2$ is dominant. 
The same effect is also demonstrated in Fig.~\ref{fig:1} right, where we show the correlation function as a function of the bin width $\Delta$.

\begin{figure}[htb!]
\begin{center}
\resizebox{0.45\textwidth}{!}{%
  \includegraphics{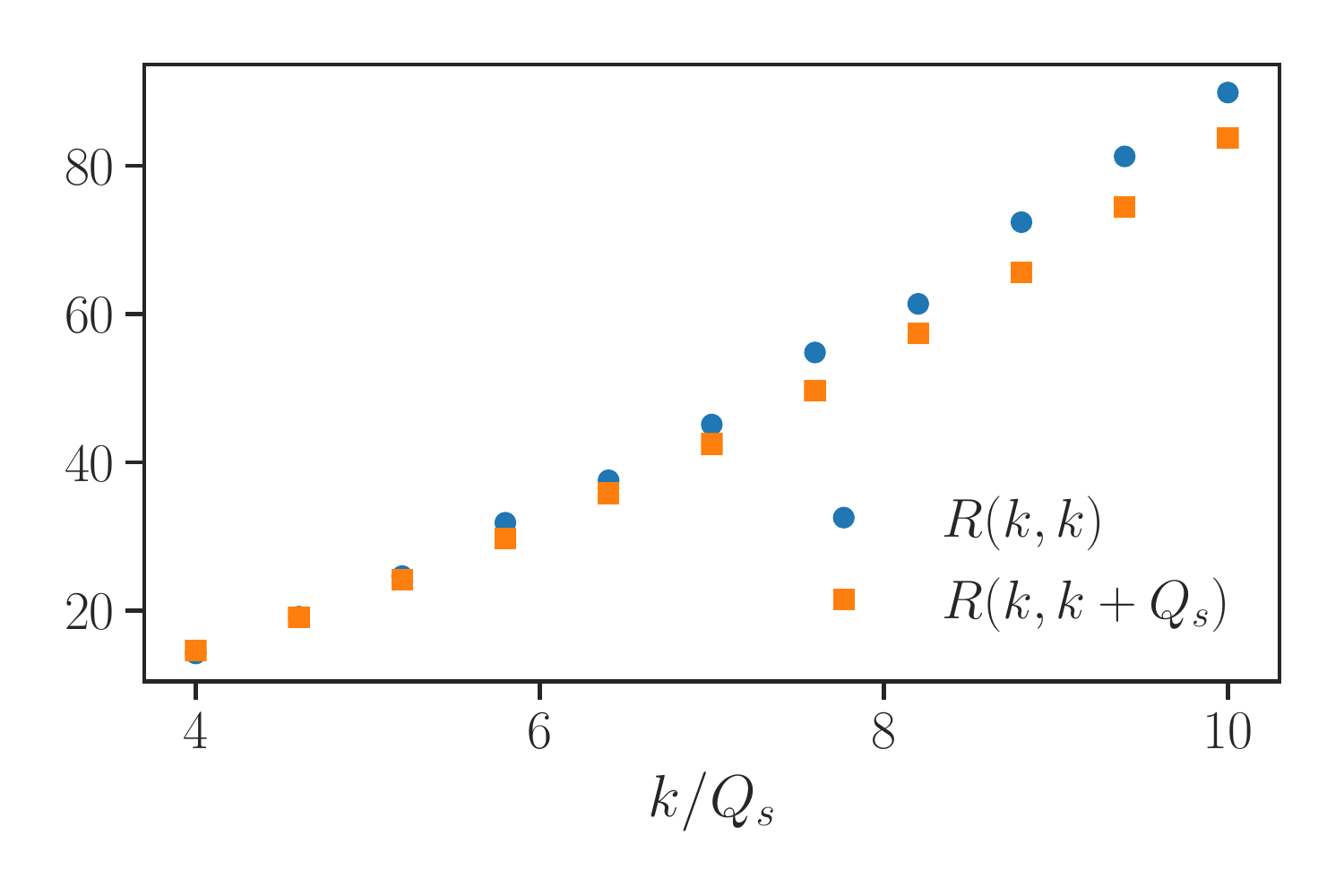}
}
\end{center}
\vskip -0.7cm
\caption{The ratio $R\equiv \langle v^2_2\overline{k^2}\rangle/\langle v^2_2N\rangle$ as a function of transverse momentum.
%Two particle correlations in $pp$ and $p$Pb collisions at the LHC measured by the ATLAS Collaboration~\cite{Aaboud:2016yar}, for different energies and particle multiplicities in the event.
}
\label{fig:5}       % Give a unique label
\end{figure}

Finally, Fig.~\ref{fig:5} shows $R\equiv \langle v^2_2\overline{k^2}\rangle/\langle v^2_2N\rangle$ as a function of transverse momentum. The correlation with transverse momentum clearly drops slower than the correlation with multiplicity.

\section{Conclusion}
\label{sec:conclu}

In this contribution we have examined~\cite{Altinoluk:2020psk} the correlation between $v_2^2$ and the total multiplicity and the squared transverse momentum per particle, in the CGC approach to multi gluon correlations in pA collisions~~\cite{Altinoluk:2018ogz}. We use several approximations, discussed in Sec.~\ref{sec:v2corr}, to make the problem tractable at an analytical level. By examining $v_2^2$ and the mentioned correlations in different bins of transverse momenta, we find that $v_2^2$ is dominated by the HBT correlation which disappears when the bins are not overlapping and solely Bose correlations contribute. Both contributions show different transverse momentum dependences. The opposite behaviour is found for the correlations, which are suppressed in those regions where the HBT contribution is dominant. The apparent strong dependence of the correlation results on small changes in transverse momenta is due to the assumption on translational invariance of the projectile that we have employed. While our results contain many approximations and we do not aim for their phenomenological application, we observe the smallness of the correlations between $v_2$ and $N$  consistent with experimental data~\cite{ATLAS:2017rtr}.

%\section*{Acknowledgements}
%Acknowledgements should follow immediately after the conclusion.
%
%% TODO: include author contributions
%\paragraph{Author contributions}
%This is optional. If desired, contributions should be succinctly described in a single short paragraph, using author initials.

% TODO: include funding information
\paragraph{Funding information}
NA has received financial support from Xunta de Galicia (Centro singular de investigaci\'on de Galicia accreditation 2019-2022), by European Union ERDF, by  the ``Mar\'{\i}a  de Maeztu" Units  of  Excellence program  MDM-2016-0692,  the Spanish Research State Agency under project FPA2017-83814-P and by the European Research Council under project ERC-2018-ADG-835105 YoctoLHC.
TA is supported by Grant No. 2018/31/D/ST2/00666 (SONA\-TA 14 - National Science Centre, Poland). 
AK is supported by the NSF Nuclear Theory grants 1614640 and 1913890.
ML was supported by the Israeli Science Foundation (ISF) grant \#1635/16.
ML and AK were also supported by the Binational Science Foundation grant  \#2018722.
%VS acknowledges support by the DOE Office of Nuclear Physics through Grant No. DE-SC0020081. VS thanks the ExtreMe Matter Institute EMMI (GSI Helmholtzzentrum f\"ur Schwerionenforschung, Darmstadt, Germany) for partial support and hospitality.
This work has been performed in the framework of COST Action CA 15213 ``Theory of hot matter and relativistic heavy-ion collisions" (THOR), MSCA RISE 823947 ``Heavy ion collisions: collectivity and precision in saturation physics''  (HI\-EIC) and has received funding from the European Un\-ion's Horizon 2020 research and innovation programme under grant agreement No. 824093. 

% TODO:
% Provide your bibliography here. You have two options:

% FIRST OPTION - write your entries here directly, following the example below, including Author(s), Title, Journal Ref. with year in parentheses at the end, followed by the DOI number.
%\begin{thebibliography}{99}
%\bibitem{1931_Bethe_ZP_71} H. A. Bethe, {\it Zur Theorie der Metalle. i. Eigenwerte und Eigenfunktionen der linearen Atomkette}, Zeit. f{\"u}r Phys. {\bf 71}, 205 (1931), \doi{10.1007\%2FBF01341708}.
%\bibitem{arXiv:1108.2700} P. Ginsparg, {\it It was twenty years ago today... }, \url{http://arxiv.org/abs/1108.2700}.
%\end{thebibliography}

% \bibliographystyle{SciPost_bibstyle} % Include this style file here only if you are not using our template
\bibliography{mybib}

\nolinenumbers

\end{document}